\documentclass{pasj00}

\newcommand{\EHM}[1]{\mbox{$\times10^{#1}$}}
\newcommand{\pmHM}[2]{\mbox{$^{#1}_{#2}$}}
\newcommand{\lumiHM}{\mbox{erg~s$^{-1}$}}
\newcommand{\fluxHM}{\mbox{erg~s$^{-1}$~cm$^{-2}$}}

\begin{document}
\SetRunningHead{H. Matsumoto et al.}{Suzaku observation of HESS~J1614$-$518}
\Received{2007/06/18}
\Accepted{2007/11/01}

\title{Discovery of Extended X-Ray emission from the unidentified
TeV source HESS~J1614$-$518 using the Suzaku Satellite}



%
 \author{%
   Hironori \textsc{Matsumoto}\altaffilmark{1},
   Hideki \textsc{Uchiyama}\altaffilmark{1},
   Makoto \textsc{Sawada}\altaffilmark{1},\\
   Takeshi G. \textsc{Tsuru}\altaffilmark{1},
   Katsuji \textsc{Koyama}\altaffilmark{1},
   Hideaki \textsc{Katagiri}\altaffilmark{2},\\
   Ryo \textsc{Yamazaki}\altaffilmark{2},
   Aya \textsc{Bamba}\altaffilmark{3},
   Kazunori \textsc{Kohri}\altaffilmark{4},
   Koji \textsc{Mori}\altaffilmark{5},\\
and\\
   Yasunobu \textsc{Uchiyama}\altaffilmark{3}}
 \altaffiltext{1}{Department of Physics, Graduate School of Science,
Kyoto University, Sakyo-ku, Kyoto 606-8502}
 \email{E-mail (HM) matumoto@cr.scphys.kyoto-u.ac.jp}
 \altaffiltext{2}{Department of Physical Science, Hiroshima University, 
Higashi-Hiroshima 739-8526}
 \altaffiltext{3}{Institute of Space and Astronautical Science,
Japan Aerospace Exploration Agency, \\
3-1-1 Yoshinodai, Sagamihara, Kanagawa 229-8510}
 \altaffiltext{4}{Physics Department, Lancaster University
LA1 4YB, UK}
\altaffiltext{5}{Department of Applied Physics, University of Miyazaki, \\
1-1 Gakuen Kibana-dai Nishi, Miyazaki 889-2192}

\maketitle

\begin{abstract}

We report the Suzaku results of HESS~J1614$-$518, which is
the brightest extended TeV gamma-ray source discovered in
the Galactic plane survey conducted using the
H.E.S.S. telescope.  We discovered three X-ray objects in
the field of view of the X-ray Imaging Spectrometer (XIS),
which were designated as Suzaku~J1614$-$5141 (src~A),
Suzaku~J1614$-$5152 (src~B), and Suzaku~J1614$-$5148
(src~C).  Src~A is an extended source located at the peak
position of HESS~J1614$-$518, and therefore it is a
plausible counterpart to HESS~J1614$-$518.  The X-ray flux
in the 2--10~keV band is 5\EHM{-13}~\fluxHM, which is an
order of magnitude smaller than the TeV flux. The photon
index is 1.7, which is smaller than the canonical value of
synchrotron emissions from high-energy electrons found in
some supernova remnants.  These findings present a challenge
to models in which the origin of the TeV emission is the
inverse Compton scattering of the cosmic microwave
background by accelerated electrons that emit X-rays via
synchrotron emission.  Src~B is located at a relatively dim
region in the TeV band image; however, its hydrogen column
density is the same as that of src~A. Therefore, src~B may
also be physically related to HESS~J1614$-$518.  Src~C is a
foreground late-type B star.  We also discovered a soft
extended X-ray emission near HESS~J1614$-$518.

\end{abstract}

\section{Introduction}

One of the striking results of the TeV $\gamma$-ray
astronomy with the H.E.S.S. Cerenkov Telescope is the
discovery of many very high-energy (VHE) objects along the
Galactic plane~\citep{Aharonian2005, Aharonian2006}. Most of
these objects are spatially extended to
$\gtrsim\timeform{0.1D}$, and could therefore be Galactic
objects, such as supernova remnants (SNR) or pulsar wind
nebulae (PWN). TeV $\gamma$-ray emissions are also detected
from a young stellar cloud, Westerlund~2
~\citep{Aharonian2007}.  These provide direct evidence that
the inner Galactic region produces high-energy particles
with energies above TeV level.  Observations of other
wavelengths, particularly X-rays, provide a better
understanding of which particles are accelerated and the
relevant dynamics. High-energy electrons predominantly emit
synchrotron X-rays in a typical interstellar magnetic field
with strength of a few micro-Gauss. Therefore, the flux
ratio of the X-rays to TeV $\gamma$-rays provides valuable
information that can distinguish whether the accelerated
particles are protons or electrons.  At present, however,
X-ray observations on these TeV sources are still limited.
\citet{Matsumoto2007} observed HESS~J1616$-$508 deeply with the Suzaku
satellite~\citep{Mitsuda2007}, in addition to analyzing the
archival data obtained from XMM-Newton. However, they found
no X-ray counterpart with a severe upper limit of
3\EHM{-13}~\fluxHM\ in the 2--10~keV band (see
also~\cite{Landi2007}). HESS~J1303$-$631~\citep{Aharonian2005b}
is the same example. On the other hand, HESS~J1804$-$216 has
some X-ray counterparts
~\citep{Landi2006,Cui2006,Bamba2007,Kargaltsev2007a};
however, their physical connection to HESS~J1804$-$216 is
still unclear.  Recently, an extended X-ray emission from
TeV~J2032$+$4130~\citep{Aharonian2002} that has a peak
position coincident with that of the $\gamma$-ray emission
was discovered as a result of a deep observation with
XMM-Newton~\citep{Horns2007}.  However, \citet{Butt2007} was
skeptical about the detection of the diffuse emission, and
argued that it can be explained by integrated emission of
faint X-ray sources detected with
Chandra~\citep{Butt2006}. Recently, an extended X-ray
emission from HESS~J1834$-$087 was discovered with
XMM-Newton~\citep{Tian2007}, which is a clear
identification.  However, there are also point-like sources
in addition to the extended emission in the spectral region
reported by \citet{Tian2007}; therefore, the physical
connection between the extended emission and
HESS~J1834$-$087 remains unclear.  Absolutely, it is
imperative to have an access to more X-ray samples with a
physical connection to the VHE objects. Therefore, we
observed HESS~J1614$-$518 (hereafter HESS~J1614) with the
Suzaku satellite, which is the brightest among the newly
discovered HESS objects~\citep{Aharonian2005,Aharonian2006}.
Two X-ray objects were found in the vicinity of HESS~J1614
using the Swift/X-Ray Telescope
(XRT)~\citep{Landi2006}. However, neither of them coincided
with the $\gamma$-ray peak of HESS~J1614; therefore their
physical connection to HESS~J1614 could not be
clarified. Since HESS~J1614 has a spatial extension of
$\sim\timeform{10'}$, high-sensitivity observations in a
hard X-ray band with low and stable backgrounds are
essential.  The Suzaku X-ray Imaging Spectrometer (XIS;
\cite{Koyama2007}), combined with the X-Ray
Telescope~(XRT; \cite{Serlemitsos2007}), satisfies these
requirements and is regarded as one of the best instruments
to search for diffuse X-ray emissions, particularly in the
hard-energy band.  Uncertainties quoted in this study are at
the 90\% confidence level, and errors on the data points in
the X-ray spectra and radial profiles are at the 1$\sigma$
confidence level, unless otherwise stated.

\section{Observations and Data Reduction}

\begin{figure*}
\begin{center}
\FigureFile(.8\textwidth,){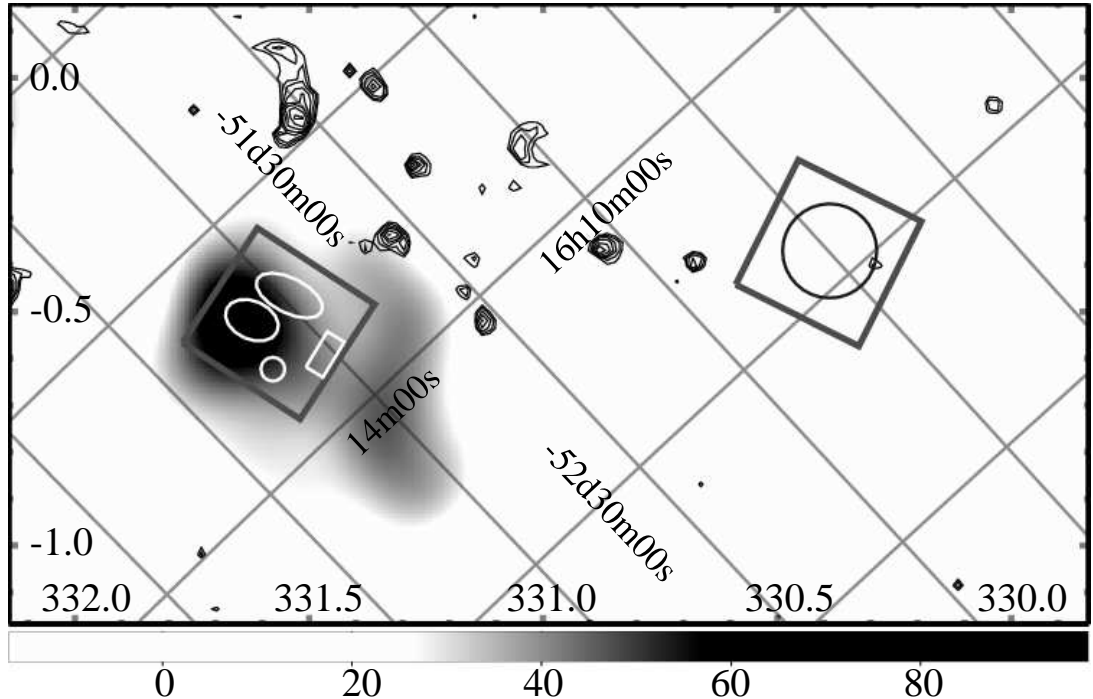}
\end{center}
\caption{Suzaku fields of view (thick boxes) overlaid 
on the H.E.S.S. smoothed excess map (same as figure~9 in
\cite{Aharonian2006}).
The scale bar below the figure represents the excess.  The
coordinates on the exterior frame are Galactic, while the
grid shows the equatorial coordinates (J2000.0).  The
contour lines denote the radio emission (843~MHz) adopted
from the Sydney University Molonglo Sky Survey (SUMSS;
\cite{Bock1999}). The contours are drawn from 0.1~Jy~beam$^{-1}$ 
on a logarithmic scale with a factor of 0.158, where the
FWHM beam size is
$\sim\timeform{0.7'}\times\timeform{0.9'}$.  Regions
outlined by thin solid lines are the ones used in our
spectral analyses (see section~\ref{sec:results}).
\label{fig:HESSvsXIS}}
\end{figure*}

The brightest part of HESS~J1614 was observed on 2006
September 16. The on-source observation was followed by a
background observation at an offset position
(figure~\ref{fig:HESSvsXIS}), where no known bright X-ray or
$\gamma$-ray source existed in the field of view.  The
latitude of the offset position is selected to be almost the
same as the position of HESS~J1614, so that the Galactic
ridge emission at the source position can be reliably
subtracted (e.g.,
\cite{Worral1982,Warwick1985,Koyama1986,Yamauchi1993,
Kaneda1997,Ebisawa2001,Ebisawa2005,Sugizaki2001,Tanaka2002,Revni2006}).
The observations are summarized in table~\ref{tbl:obs}.

\begin{table*}
\caption{Log of Suzaku observations.
\label{tbl:obs}}
\begin{center}
\begin{tabular}{cccccc} \hline \hline
Name&OBSID &\multicolumn{2}{c}{Pointing direction} 
&Observation start (UT)&Effective exposure (ks) \\ 
&&$l$&$b$&&\\ \hline
HESS~J1614 &501042010&\timeform{331.5717D}&\timeform{-0.5274D}&2006/09/15 16:00&44.5\\
offset&501043010&\timeform{330.3961D}&\timeform{-0.3760D}&2006/09/16 11:02&53.1\\ 
\hline
\end{tabular}
\end{center}
\end{table*}

The observations were performed with the four CCD cameras
(XIS) located at the focal planes of four XRTs. One of the
XIS sensors (XIS1) has a back-illuminated (BI) CCD, while
the other three sensors (XIS0, 2, and 3) utilize
front-illuminated (FI) CCDs. The XIS was operated in the
normal clocking mode with no charge injection technique.
Although the non-imaging Hard X-ray Detector (HXD;
\cite{Kokubun2007,Takahashi2007}) was also available, 
in this study, we focus on the XIS data analysis because our
main objective is searching for X-ray counterparts of
HESS~J1614 using the imaging capability.

Data were reduced and analyzed using the processed data of
version 1.2.2.3\footnote{ See
http://www.astro.isas.jaxa.jp/suzaku/process/history/v1223.html},
the HEADAS software version 6.2, and a calibration database
(CALDB) released on 2007 April 9.  All data affected by the
South Atlantic Anomaly and/or telemetry saturation were
excluded. We also excluded the data obtained with low
elevation angles from the Earth rim that were below
$<\timeform{5D}$
\footnote{We did not screen the data with the elevation 
angles from the bright Earth rim in order to maximize the
statistics, although the standard data processing done at
ISAS/JAXA or NASA/GSFC usually use this screening. The
bright Earth may contaminate the soft-band spectra. In our
case, however, we confirmed that the standard screening
reduced the effective exposures but did not affect the XIS
spectra.  For example, excluding the data with low elevation
angles from the bright Earth rim of $<\timeform{20D}$
changed the counting rate of XIS1 in the 0.2--2~keV band
from 0.1722~cts~s$^{-1}$ to 0.1726~cts~s$^{-1}$, while the
exposure time decreased from 44.5ks to 40.3ks.}.  Finally,
we removed the hot and flickering pixels. After these data
screenings, the effective exposures were 44.5~ks and 53.1~ks
on the HESS~J1614 and the offset background, respectively.

We checked the energy scale of each XIS sensor using the
calibration source (\atom{Fe}{}{55}) and confirmed that the
observed center energies of K$\alpha$ and K$\beta$ lines
were in good agreement with the expected values (5.895 and
and 6.490~keV) within the nominal calibration uncertainty of
0.2\%~\citep{Koyama2007}.

\section{Analysis and Results
\label{sec:results}}

\subsection{XIS image
\label{subsec:image}}

We extracted XIS images from each sensor using the screened
data for the soft- and hard-energy bands. For the FI
sensors, the soft- and hard-bands are defined as
0.4--3.0~keV and 3.0--10.0~keV, respectively, while those
for the BI sensor are defined as 0.3--3.0 and 3.0--7.0~keV,
respectively.  We excluded the corners of the CCD chips
illuminated by the
\atom{Fe}{}{55} calibration sources. Images of the non-X-ray
background (NXB) were obtained from the night Earth data
provided by the XIS team~\citep{Tawa2007} and subtracted
from the HESS~J1614 images. Then, the soft and hard images
were divided by flat sky images simulated at 1.49 and
6.0~keV using the XRT+XIS simulator {\tt
xissim}~\citep{Ishisaki2007} for vignetting corrections.
The images from the three FI sensors were summed and
re-binned by a factor of 8.

\begin{figure*}
\begin{center}
\FigureFile(.9\textwidth,){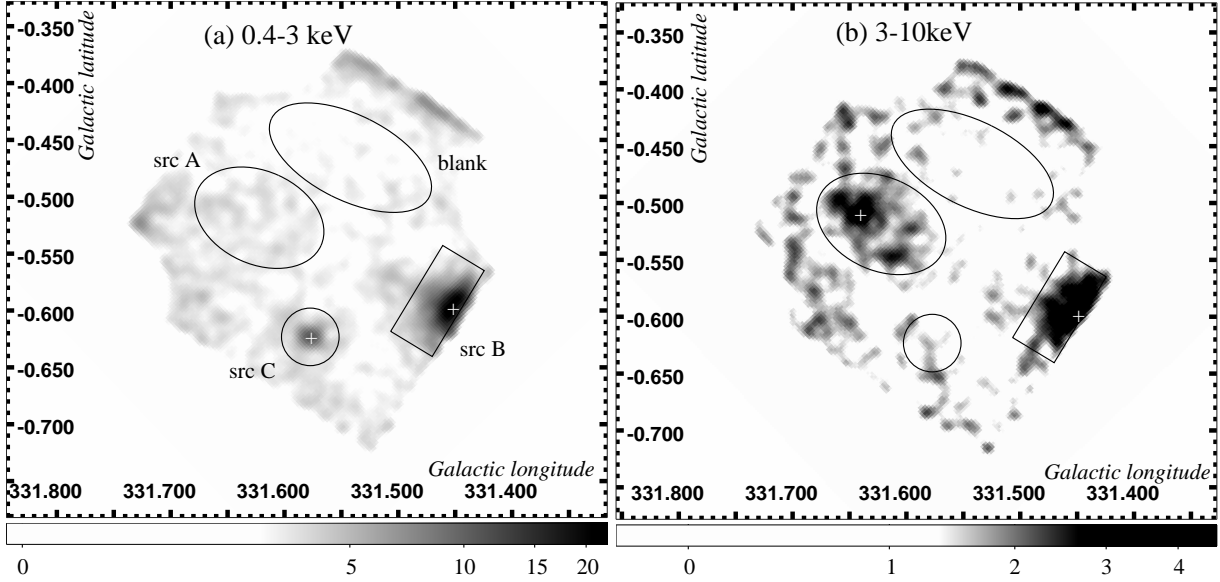}
\end{center}
\caption{Suzaku XIS FI (XIS0+2+3) images of the HESS~J1614 region in 
the Galactic coordinates: (a) 0.4--3~keV and (b) 3--10~keV
band. The images were smoothed using a Gaussian function
with a sigma of \timeform{0.28'}. Vignetting correction was
applied after subtracting NXB as described in the text.
Regions represented with solid lines are used in our
spectral analyses (see section~\ref{subsec:spec}).  Plus
marks represent the peak position of each source.
\label{fig:XISimage}}
\end{figure*}

The XIS FI images of the HESS~J1614 region shown in
figure~\ref{fig:XISimage} were smoothed using a Gaussian
function with a sigma of $\timeform{0.28'}$. The BI images
were essentially the same except for the poorer statistics.
In the hard-band image, an extended object with a peak
position of $(l, b)=(\timeform{331.64D},
\timeform{-0.51D})$\footnote{$(\alpha, \delta)_{\rm
J2000.0}=(\timeform{16h14m34s}, \timeform{-51D41'00''})$}
was found, and was designated as Suzaku~J1614$-$5141
(src~A).  The position uncertainty was
\timeform{0.8'}.  Src~A was not conspicuous in the soft-band
image. Another bright X-ray object, found at $(l,
b)=(\timeform{331.45D},
\timeform{-0.60D})$
\footnote{$(\alpha, \delta)_{\rm J2000.0}=(\timeform{16h14m06s}, \timeform{-51D52'38''})$}, was also bright
in the soft X-ray band. The position uncertainty was
\timeform{0.8'}.  We designated this object as
Suzaku~J1614$-$5152 (src~B).  In the soft-band image, an
X-ray source was found at $(l, b)=(\timeform{331.58D},
\timeform{-0.62D})$
\footnote{$(\alpha, \delta)_{\rm J2000.0}=(\timeform{16h14m47s}, \timeform{-51D48'36''})$}
and was designated as Suzaku~J1614$-$5148 (src~C). The
position uncertainty was \timeform{0.6'}.

\begin{figure}
\begin{center}
\FigureFile(.45\textwidth,){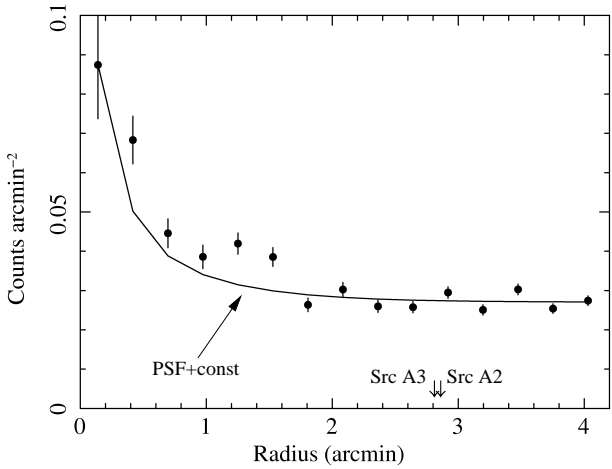}
\end{center}
\caption{
Radial profile of src~A extracted from the hard-band image
of the XIS FI sensor (XIS0+2+3). The solid line represents
the XIS PSF profile with a constant component. The downward
arrows represent the positions of src~A2 and A3 defined in
section~\ref{sssec:srcA} (see figure~\ref{fig:srcA123}).
\label{fig:srcA_radpro}}
\end{figure}

We obtained the radial profiles and compared with a
point-spread function (PSF).  As for the PSF, we obtained
the radial profiles using the SS~Cyg data observed on 2005
November 2 (OBSID$=$400006010), which is the verification
phase data for the imaging capability of the
XRT~\citep{Serlemitsos2007}. Since the energy dependence of
the PSF is almost negligible~\citep{Serlemitsos2007}, it was
extracted from the 0.4--10~keV band.
Figure~\ref{fig:srcA_radpro} shows the radial profile of
src~A in the hard-energy band.  The profile cannot be fitted
with the PSF plus a constant component model ($\chi^2/{\rm
d.o.f.}=45.7/13$), and therefore src~A should be an extended
source. Since src~B is located near the edge of the XIS
field of view, and therefore the PSF is largely distorted,
it is impossible to study the radial profile of src~B in
detail.  However, the hard-band image suggested that src~B
is extended.  The radial profile of src~C is consistent with
the PSF plus a constant component ($\chi^2/{\rm
d.o.f.}=37.1/26$), supporting the assumption of a point
source.

\subsection{XIS spectrum
\label{subsec:spec}}

\subsubsection{Background spectra: blank and offset positions}

Since the source position is on the Galactic ridge, the
local Galactic emission (the ridge emission), as well as the
NXB and cosmic X-ray background, cannot be ignored.  To
account for the background, we compared the spectrum from
the blank region near the sources (see the solid line in
figure~\ref{fig:XISimage}) with that from the offset
position.  The spectra of the offset position correspond to
a circular region of \timeform{6'} radius at the center of
the field of view.

\begin{figure*}[t]
\begin{center}
(a)\FigureFile(0.45\textwidth,){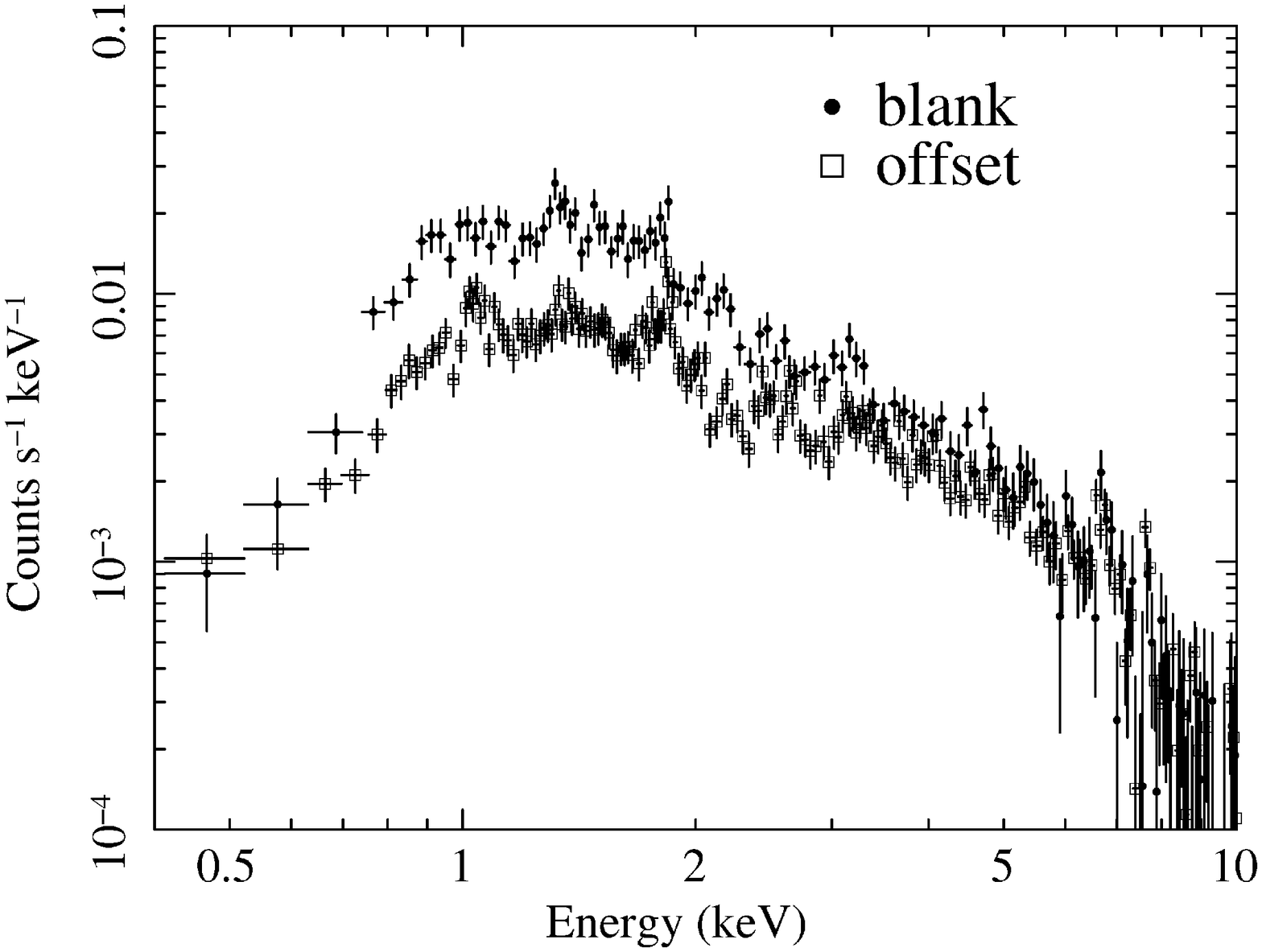}
(b)\FigureFile(0.45\textwidth,){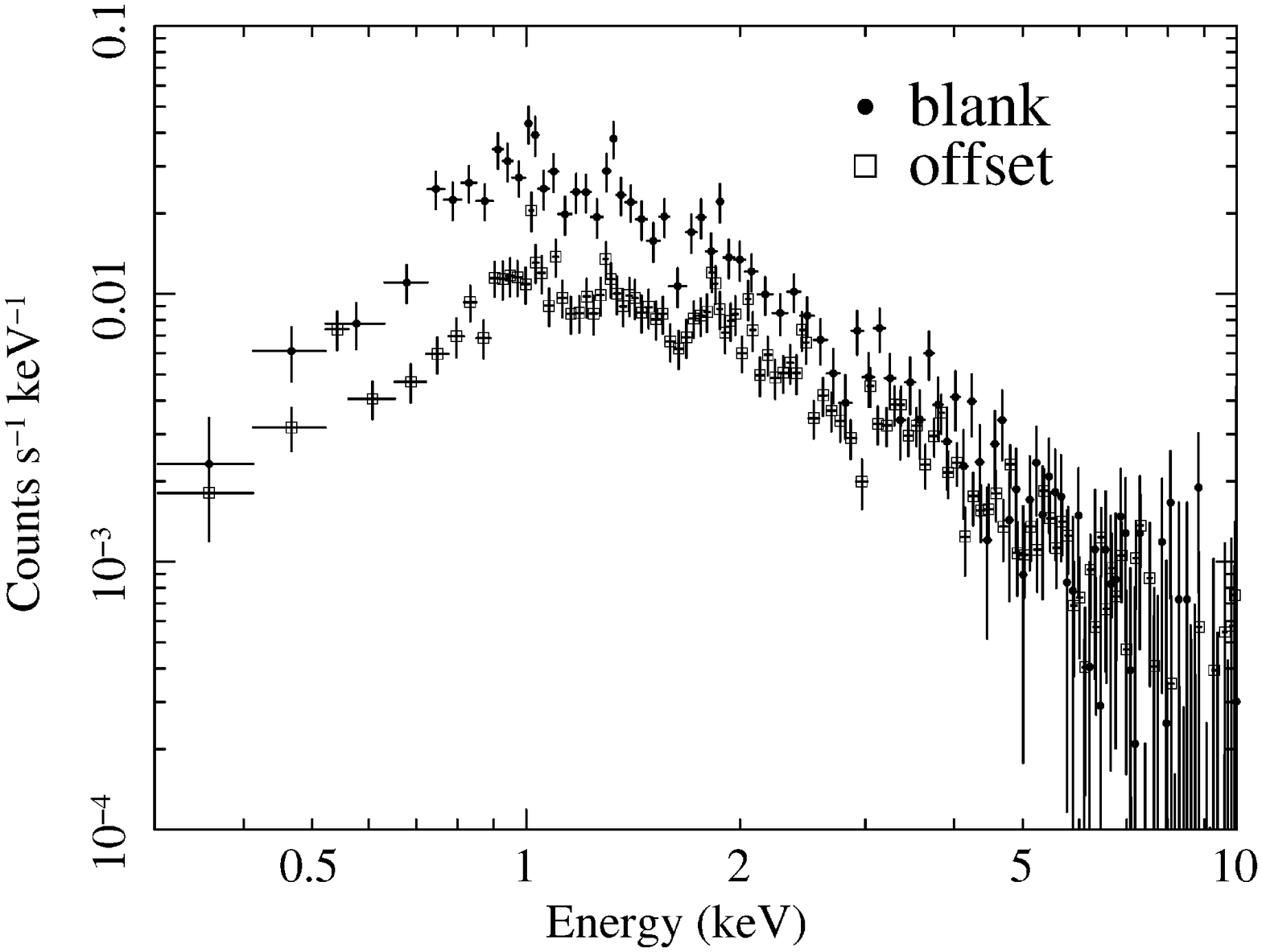}
\end{center}
\caption{
XIS spectra of the blank and offset regions: (a) FI spectra
(XIS0+2+3) and (b) BI (XIS1) spectra. The offset spectra
were normalized to the same area as that of the blank
region. The NXBs were subtracted as described in the text.
\label{fig:blank_offset}}
\end{figure*}

For the most accurate NXB estimation, we sorted the night
Earth data so that the cut-off rigidity distribution was the
same as that of the HESS~J1614 and offset observations.
Since the NXB depends on a location on the
detector~\citep{Yamaguchi2006}, we extracted the NXB spectra
from the same regions used to determine the blank and offset
regions in detector coordinates (DETX/Y). The NXB spectra
thus obtained were then subtracted, and the resulting
spectra of the blank and offset regions are shown in
figure~\ref{fig:blank_offset}. The offset spectra were
normalized to the same area as that of the blank region.  We
can clearly observe that the spectra of the blank region
show an enhancement less than 3.0~keV.

\begin{figure}
\begin{center}
\FigureFile(.45\textwidth,){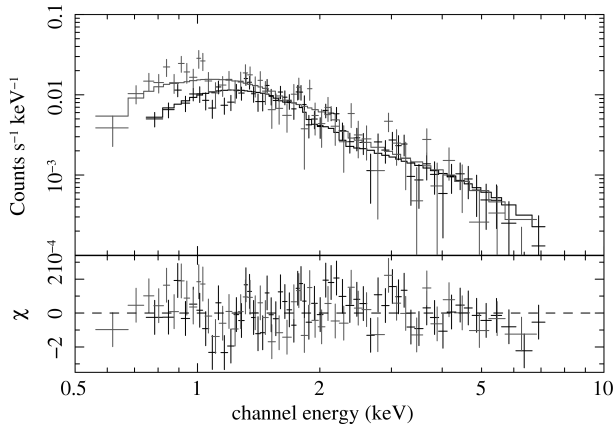}
\end{center}
\caption{Spectra of the blank region and the best-fit power-law model.
The spectra of the offset region are subtracted as
background.  Black and gray lines represent the data and model for
the XIS FI (XIS0+2+3) and BI (XIS1), respectively.
\label{fig:blank_spec}}
\end{figure}

The excess spectra of the blank region after subtracting the
data of the offset position are shown in
figure~\ref{fig:blank_spec}.  We fitted these spectra with
an absorbed power-law model, and the cross sections of
photoelectric absorption were obtained
from~\citet{Morrison1983}. We obtained detector responses
(RMF) and telescope responses (ARF) using both the {\tt
xisrmfgen} and {\tt xissimarfgen}
software~\citep{Ishisaki2007}.  In this case, a
hydrogen-equivalent column density ($N_{\rm H}$), a photon
index ($\Gamma$), and normalization are free parameters. The
best-fit parameters are summarized in
table~\ref{tbl:spec_results}. Although the data were fitted
acceptably with the power-law model, we can see some
indications of emission lines in the residual, particularly
at $\sim$1.3~keV. When we added a Gaussian line model with
$\sigma=0$ (i.e., the delta function) to the power-law
model, the peak energy became 1.33\pmHM{+0.04}{-0.04}~keV
and the equivalent width was 24.0 ($<$53.1)~eV, although the
F-test does not require the use of a Gaussian model. These
results suggest the emission line of MgXI and a thermal
spectrum. Therefore we tried to fit the spectra with a
thin-thermal plasma model (the APEC model;
\cite{Smith2001}), where the solar abundance of elements 
is obtained from~\citet{Anders1989}. The free parameters are
plasma temperature ($kT$), metal abundance, $N_{\rm H}$, and
normalization. The spectra were fitted acceptably with the
plasma model, as summarized in
table~\ref{tbl:spec_results}. However, the metal abundance
is almost zero.

Since the blank region is in a dim area of both the soft-
and hard-energy bands (figure~\ref{fig:XISimage}), the soft
excess must also exist in the regions of src~A, B, and C.
As for the source spectra, we concern the excess flux over
this soft excess emission, and therefore used the blank
region as background in the following analyses of the source
spectra.

\subsubsection{Extraction of the source spectra}

We represented the source regions of src~A, B, and C as
solid lines in figure~\ref{fig:XISimage}. We extracted light
curves corresponding to the sources in the soft- and
hard-energy bands from these regions and observed that none
of them exhibited clear time variability.  We extracted the
XIS spectra of src~A, B, and C by subtracting the NXB
spectra, in the same manner as in the previous section. We
then combined the NXB subtracted spectra obtained from three
FI sensors.

\subsubsection{Src~A
\label{sssec:srcA}}

\begin{figure}
\begin{center}
\FigureFile(.45\textwidth,){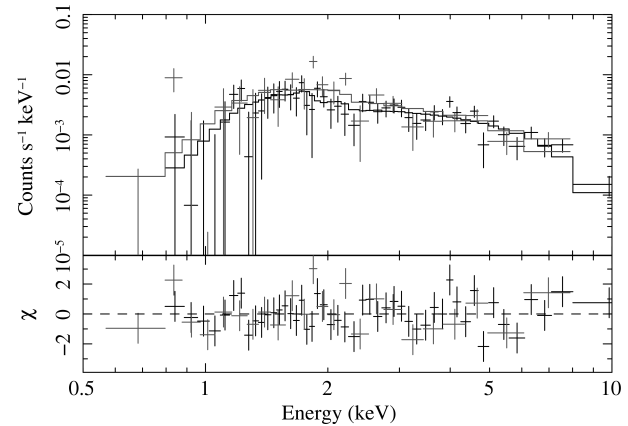}
\end{center}
\caption{
XIS spectra of src~A and the best-fit power-law model. Black
and gray lines represent the data and the model for the XIS
FI (XIS0+2+3) and BI (XIS1), respectively.
\label{fig:srcAall_spec}}
\end{figure}

The spectra of src~A are shown in
figure~\ref{fig:srcAall_spec}. Although the statistics are
limited, no clear emission line was observed. We therefore
fitted the spectra with the absorbed power-law model and
obtained acceptable results. The best-fit parameters are
summarized in table~\ref{tbl:spec_results}. In addition, we
also tried to fit the spectra with the thermal plasma
model. The best-fit parameters were $N_{\rm
H}=1.04\pmHM{+0.34}{-0.36}\EHM{22}$~cm$^{-2}$,
$kT=10.0\pmHM{+26.0}{-4.2}$~keV, and an abundance of
0.18($<0.93$)~solar with $\chi^2/{\rm
d.o.f.}=80.01/67$. From a statistical viewpoint, neither of
the models is rejected at a confidence level of 90~\%.
However, the thermal plasma model yields an uncomfortably
high temperature and an extremely low abundance, and
therefore is not practicable. We therefore adopted the
power-law model in all subsequent analyses and discussions.

\begin{figure}
\begin{center}
(a)\FigureFile(.45\textwidth,){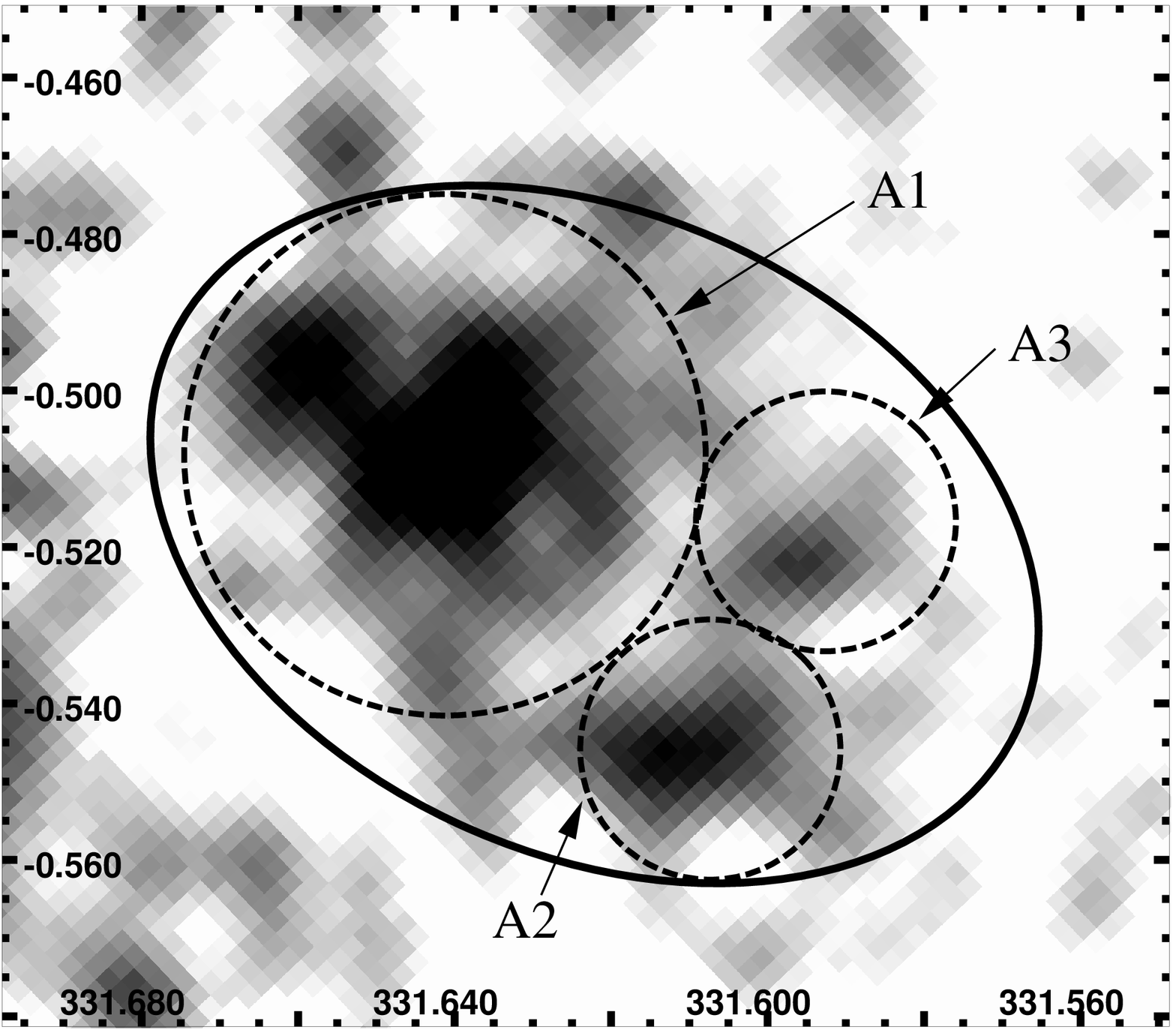}
(b)\FigureFile(.45\textwidth,){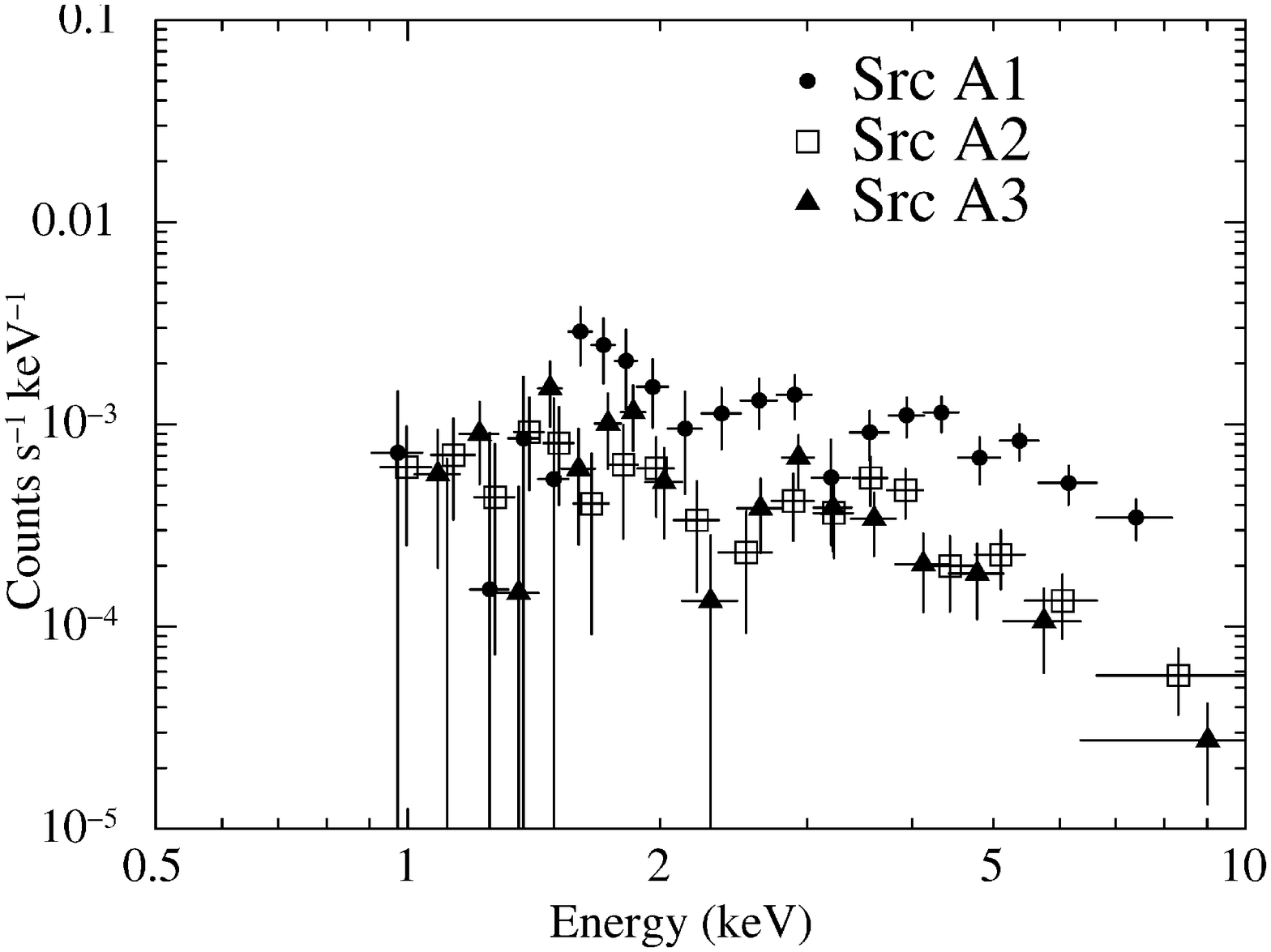}
\end{center}
\caption{(a) Close-up view of src~A in the XIS FI hard-band image 
(figure~\ref{fig:XISimage}). A large ellipse shows the
spectral region for src~A. Dotted circles designated as A1,
A2, and A3 are the regions we studied to check for spatial
variation of the spectra in src~A. (b) Spectra extracted
from regions A1, A2, and A3, as shown in
figure~\ref{fig:srcA123}(a).  For clarity, only the spectra
obtained by the FI sensors (XIS0+2+3) are shown.
\label{fig:srcA123}}
\end{figure}

Figure~\ref{fig:srcA123}(a) is a close-up view of src~A in
the XIS FI hard-band image (figure~\ref{fig:XISimage}). One
could argue that src~A is not a single extended source, but
instead consists of multiple objects. To verify this
possibility, we defined three circular regions designated as
A1, A2, and A3 in the figure, and extracted the
corresponding spectra. The radii of the source regions are
\timeform{2'},
\timeform{1'}, and
\timeform{1'} for A1, A2, and A3, respectively. 
However, the statistics are poor and with the exception of
the corresponding normalizations, no clear difference
between these sources was observed
(figure~\ref{fig:srcA123}(b)).  The spectra are described by
a power-law model with a best-fit $\chi^2/{\rm d.o.f.}$ of
25.72/31, 13.35/23, and 23.03/23, for A1, A2, and A3,
respectively. The observed fluxes in the 0.5--10.0~keV band
are 2.9\EHM{-13}~\fluxHM, 1.1\EHM{-13}~\fluxHM, and
0.95\EHM{-13}~\fluxHM, for A1, A2, and A3, respectively. The
best-fit parameters are summarized as follows: $N_{\rm
H}=1.40\pmHM{+0.86}{-0.67}\EHM{22}$~cm$^{-2}$ and
$\Gamma=1.41\pmHM{+0.46}{-0.42}$ for the A1 region, $N_{\rm
H}=0.66\pmHM{+0.91}{-0.57}\EHM{22}$~cm$^{-2}$ and
$\Gamma=1.39\pmHM{+0.57}{-0.49}$ for the A2 region, and
$N_{\rm H}=1.47\pmHM{+0.99}{-0.70}\EHM{22}$~cm$^{-2}$ and
$\Gamma=2.26\pmHM{+0.72}{-0.58}$ for the A3 region.  All
these parameters are consistent with the best-fit values of
src~A presented in table~\ref{tbl:spec_results}, which
supports the hypothesis that src~A1, A2, and A3 are
substructures of src~A.

\subsubsection{Src~B}

\begin{figure}
\begin{center}
\FigureFile(.45\textwidth,){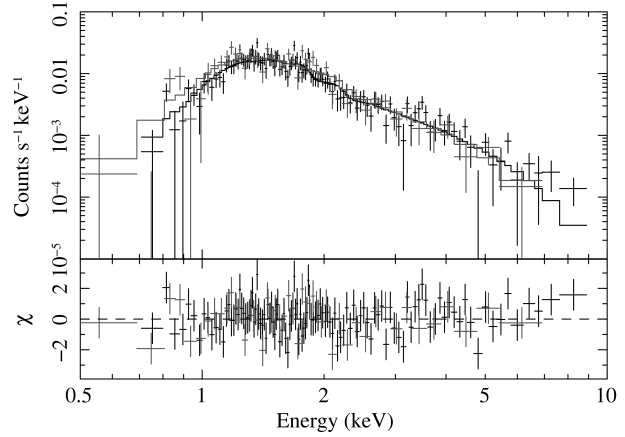}
\end{center}
\caption{
XIS spectra of src~B, shown with the best-fit power-law
model. Black and gray lines represent the data and the model
for the XIS FI (XIS0+2+3) and BI (XIS1), respectively
\label{fig:srcB_spec}}
\end{figure}

The spectra of src~B are shown in
figure~\ref{fig:srcB_spec}.  We fitted the power-law model
with the best-fit parameters as listed in
table~\ref{tbl:spec_results}. Although no emission line was
observed in the spectra, we also tried fitting a thermal
plasma model. The best-fit $\chi^2$ value of 196.13 for 159
degrees of freedom rejected the validity of a thermal plasma
model at a confidence level of 97\%.  Nevertheless, for the
sake of comparison, following are the best-fit parameters:
$N_{\rm H}=0.84\pmHM{+0.09}{-0.11}\EHM{22}$~cm$^{-2}$,
$kT=1.28\pmHM{+0.21}{-0.13}$~keV, and a metal abundance of
zero ($<0.019$)~solar.

Src~B is at the edge of the field of view. According to a
simulation performed using {\tt xissim}, the probability of
detecting incident photons reduces by a factor of $\sim$0.7
because some photons fall outside of the field of
view. Furthermore, considering the fluctuation of the
pointing direction due to the variation of the relative
alignment between the XRT and the Attitude and Orbit
Controlling System (AOCS; \cite{Serlemitsos2007}), even more
photons can escape from the field of view. Consequently, all
the obtained parameters, particularly flux, could have large
systematic errors.

\subsubsection{Src~C}

\begin{figure}
\begin{center}
\FigureFile(.45\textwidth,){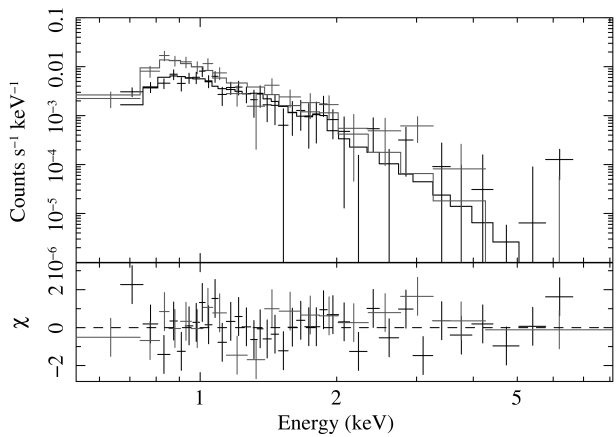}
\end{center}
\caption{
XIS spectra of src~C, shown with the best-fit thermal plasma
model. Black and gray lines represent the data and the model
for the XIS FI (XIS0+2+3) and BI (XIS1), respectively
\label{fig:srcC_spec}}
\end{figure}

We found that the spectra of src~C can be fitted with the
thermal plasma model, as shown in
figure~\ref{fig:srcC_spec}.  The best-fit parameters are
listed in table~\ref{tbl:spec_results}. Although the spectra
can also be fitted with a power-law model having best-fit
values as $N_{\rm
H}=1.01\pmHM{+0.53}{-0.36}\EHM{22}$~cm$^{-2}$, and
$\Gamma=8.63\pmHM{+3.29}{-2.25}$ with $\chi^2/{\rm
d.o.f.}=63.76/52$, the obtained photon index is unreasonably
large. Therefore, the thermal plasma model seems more
suitable for practical use.

\begin{table*}
\caption{Best-fit results of the XIS spectra.
\label{tbl:spec_results}}
\begin{center}
\begin{tabular}{lccccc}\\ \hline \hline
&\multicolumn{2}{c}{blank}&src~A&src~B&src~C\\ \hline 
BGD$^\dag$&offset&offset&blank&blank&blank\\
Model$^\ddag$&PL&APEC&PL&PL&APEC\\ 
$N_{\rm H}$ ($10^{22}$cm$^{-2}$)
&0.24\pmHM{+0.08}{-0.07}&0.10\pmHM{+0.05}{-0.04}
&1.21\pmHM{+0.50}{-0.41}&1.24\pmHM{+0.14}{-0.13}&0.28\pmHM{+0.19}{-0.19}\\
$\Gamma$/$kT$ (keV)
&2.38\pmHM{+0.19}{-0.17}&2.72\pmHM{+0.50}{-0.43}
&1.73\pmHM{+0.33}{-0.30}&3.60\pmHM{+0.24}{-0.22}&0.63\pmHM{+0.15}{-0.06}\\
Abundance (solar) 
&---&0.00($<$0.029)
&---&---&0.18\pmHM{+1.10}{-0.10}\\ 
$F^{\rm obs}_{\rm 0.5-10keV}$$^{\P}$ 
&0.10&0.096&5.4&5.1&0.86\\ 
$F^{\rm abscor}_{\rm 0.5-10keV}$$^{\S}$
&0.15&0.11&8.4&35.8&2.0\\ 
$F^{\rm obs}_{\rm 2-10keV}$$^{\P}$ 
&0.058&0.050&4.8&3.0&0.071\\
$F^{\rm abscor}_{\rm 2-10keV}$$^{\S}$ 
&0.059&0.051&5.3&3.6&0.076\\
$\chi^2/{\rm d.o.f.}$ 
&121.45/103&111.37/102&78.38/68&172.65/160&42.45/51\\ \hline
\end{tabular}
\end{center}
$\dag$ Background spectrum used for the analysis.\\
$\ddag$ Model used for the spectrum fitting: ``PL'' is a power-law model, and ``APEC'' is a thin thermal plasma model.\\
$\P$ Observed flux in the 0.5--10 keV or 2--10keV band in the unit 
of $10^{-13}$~\fluxHM. For the blank region, values are normalized to
a region of 1 arcmin$^2$ (i.e., \fluxHM~arcmin$^{-2}$).\\
$\S$ Absorption corrected flux in the 0.5--10 keV or 2--10keV 
band in the unit of $10^{-13}$~\fluxHM. 
For the blank region, values are normalized to
a region of 1 arcmin$^2$ (i.e., \fluxHM~arcmin$^{-2}$).\\
\end{table*}

\section{Discussion}

\subsection{Src~A}

Src~A is the closest source to the peak position of
HESS~J1614, and both the X-ray and TeV $\gamma$-ray
emissions are extended. Therefore, src~A is a plausible
X-ray counterpart to HESS~J1614. The X-ray spectrum is
described using the power-law model with $\Gamma=1.73$, and
the absorption-corrected flux in the 2--10~keV band is
5.3\EHM{-13}\fluxHM, while the $\gamma$-ray spectrum yields
$\Gamma$=2.46 and the flux in the 1--10~TeV band from a
circular region of \timeform{0.4D} is estimated to be
1.8\EHM{-11}~\fluxHM~\citep{Aharonian2006}.  Therefore, the
flux ratio $F(1-10~{\rm TeV})/F(2-10~{\rm keV})$ is
$\sim$34, which is one of the largest values observed among
extended VHE objects (see \cite{Matsumoto2007} and
references therein).

We first examined the difference between the peak positions
of src~A and HESS~J1614. Since the peak position of TeV
$\gamma$-ray is not presented in the study by
\citet{Aharonian2006}, we only worked in the bright region of
figure~\ref{fig:HESSvsXIS} and fitted it with a Gaussian
function. The peak position thus estimated is $(l,
b)=(\timeform{331.65D},\timeform{-0.52D})$\footnote{$(\alpha,
\delta)_{\rm
J2000.0}=(\timeform{16h14m41s},\timeform{-51D41'03''})$},
which is $\sim$\timeform{0.8'} away from the X-ray peak of
src~A. Considering that the best H.E.S.S. angular resolution
is $\sim$\timeform{5'}~\citep{Aharonian2006}, and from the
position uncertainty of src~A ($\sim$\timeform{0.8'}), we
can conclude that src~A spatially coincides with the peak of
HESS~J1614.

If src~A is physically associated with HESS~J1614, the
best-fit column density ($1.2\EHM{22}$~cm$^{-2}$) suggests
that the distance to HESS~J1614 is roughly $\sim$10~kpc,
since the best-fit value is approximately equal to half of
the total Galactic HI column density towards the HESS~J1614
region ($\sim2.2\EHM{22}$~cm$^{-2}$;
\cite{Dickey1990}). Assuming a distance of 10~kpc, the luminosity
of src~A in the 2--10~keV band is 6\EHM{33}~\lumiHM.

This region was also observed with the Swift
XRT~\citep{Landi2006}; however, no source was found at the
position of src~A. This is probably due to the limited
exposure time ($\sim1700$~s), and/or the small effective
area of the Swift XRT. Furthermore, src~A is near the edge
of the field of view of the Swift XRT, which makes the
detection more difficult.  No counterparts in the other
wavelengths are found in literatures.

\begin{figure}
\begin{center}
\FigureFile(.45\textwidth,){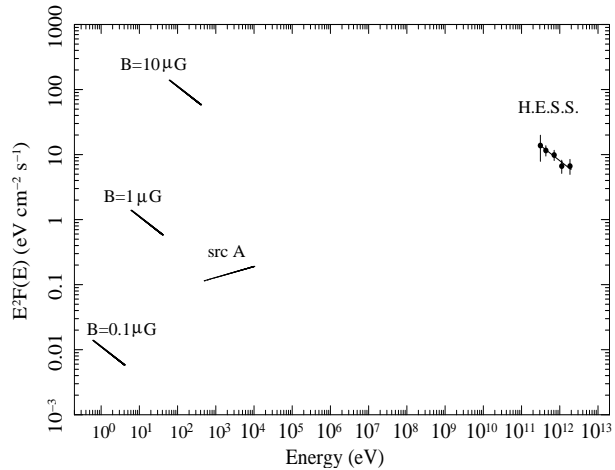}
\end{center}
\caption{
Spectral energy distribution of src~A and HESS~J1614 from
the X-ray to TeV $\gamma$-ray bands. The synchrotron
radiation from accelerated electrons, which kick the 3K
background up to the TeV energy range, is plotted toward the
left for three different values of the magnetic field
intensity. The regions used to estimate the SED in the TeV
$\gamma$-ray and X-ray bands are different.
\label{fig:SED}}
\end{figure}

If the origin of the TeV emission is the inverse Compton
scattering of the cosmic microwave background by high-energy
electrons of a single population, the large flux ratio
$F(1-10~{\rm TeV})/F(2-10~{\rm keV}) \sim 34$ requires a low
magnetic field. The spectral energy distribution (SED) of
src~A and HESS~J1614 are plotted in figure~\ref{fig:SED}. In
the same figure, we also plotted the estimated synchrotron
flux of the high-energy electrons in the magnetic fields
with intensities $B=$0.1, 1, and 10~$\mu$Gauss.  The low
X-ray flux of src~A requires a magnetic field of
$B\lesssim1\mu{\rm G}$, which is smaller than the typical
interstellar magnetic field intensity of a few
micro-Gauss. This is similar to what is observed in
HESS~J1616$-$508~\citep{Matsumoto2007}; however, but in that
case, a possibility of a magnetic field of a few micro-Gauss
and a strong cut-off in the electron energy distribution
greater than $\sim10^{14}$~eV cannot be excluded. However,
Src~A has a small spectral index, which does not favor this
possibility.  

On the other hand, interstellar radiations whose wavelengths
are shorter than those of the CMB (e.g., far-infrared) also
have large energy density~\citep{Moskalenko2006}.  If the
seeds of the inverse Compton are such photons, the electrons
produce synchrotron emission at wavelengths longer than
X-rays in regions with a magnetic field of a few
micro-Gauss, and therefore our Suzaku results would have no
constraints on the magnetic field intensity.

Since src~A is extended, it may be an SNR. However, the
observed X-rays are probably not due to ordinary synchrotron
emission, because the photon index ($\sim$1.7) is smaller
than the typical values of non-thermal SNRs such as SN1006
($\sim$2.5--3.0; \cite{Bamba2005}).  The X-ray spectrum of
the SNR $\gamma$~Cygni also has a small photon index of
0.8--1.5~\citep{Uchiyama2002}, which can be considered as a
non-thermal bremsstrahlung from accelerated electrons; the
energy distribution of the accelerated electrons below the
Coulomb break is flattened due to the Coulomb interactions
with ambient molecular clouds~\citep{Uchiyama2002}. In that
scenario, a photon index less than 1.5 is expected for
viable acceleration mechanisms. Since the photon index of
src~A is marginally consistent with this criterion, the
observations of src~A could be explained with this
scenario. In addition, undetected molecular clouds
responsible for the bremsstrahlung may
exist. \citet{Yamazaki2006} suggested that the TeV-to-X-ray
flux ratio of old SNRs with an age of $\sim 10^5$~years can
be very large, in some cases more than 100. The small photon
index of src~A may support this scenario, where X-ray
emission is dominated by synchrotron emission from secondary
electrons that may have a smaller index than that of the
primary electrons.  However, no radio counterpart has been
observed in the HESS~J1614 region, which could pose a severe
problem for the SNR scenarios.  For example, there is no
enhancement in the 843~MHz band
(figure~\ref{fig:HESSvsXIS}), where the rms noise level is
$\sim$2~mJy~arcmin$^{-2}$~\citep{Bock1999}.

Massive stars can be the origin of the extended TeV
$\gamma$-ray emission; \citet{Torres2007} proposed that
stellar winds can produce hadronic $\gamma$-rays by
interacting with cosmic rays. \citet{Ancho2007} argue that
TeV $\gamma$-rays are the result of Lorentz-boosted MeV
$\gamma$-rays, which are emitted due to the de-excitation of
daughter nuclei resulting from collisions between
high-energy nuclei (the cosmic rays) and UV photons from the
massive stars. In either case, X-rays may originate from
stellar winds from the OB stars.  These models have been
proposed in order to explain the origin of the unidentified
TeV source, TeV~J2032$+$4130.  Although the X-ray photon
index of TeV~J2032$+$4130 is similar to that of src~A, the
flux ratio $F(1-10~{\rm TeV})/F(2-10~{\rm keV})$ ($\sim$
2--3) is much smaller than that of src~A~\citep{Horns2007}.
Furthermore, no OB stars have been observed around HESS1614.

Another possibility is a PWN; some PWNe have a small photon
index~(\cite{Fleishman2007} and references therein). The
luminosity, assuming the distance of 10~kpc, is not unusual
for X-ray PWNe~\citep{Cheng2004}. A pulsar driving the
nebula may be hidden in the middle of src~A, or src~B itself
could be the pulsar.  However, the flux ratio $F(1-10~{\rm
TeV})/F(2-10~{\rm keV})$ is unusually large compared with
the typical PWNe; the ratios of Crab and RCW89 are
2.6\EHM{-3} and 0.41,
respectively~\citep{Willingale2001,Aharonian2004,Aharonian2005c,DeLaney2006}.
In addition, the lack of radio emissions may be difficult to
explain by assuming a PWN origin.

\subsection{Src~B}

The best-fit value of the column density suggests that the
distance to src~B from us is similar to that to src~A. There
is an optical~\citep{Monet1999} and an infrared
counterpart~\citep{Skrutskie2006}, 2MASS~J16140610-5152264,
at $(l, b)=(\timeform{331.4579D},
\timeform{-00.5973D})$, whose color also suggests 
extinction~\citep{Landi2006}. Assuming a distance of 10~kpc,
the luminosity of src~B in the 2--10 keV band is
4\EHM{33}~\lumiHM.

Src~B was also detected with the
Swift XRT~\citep{Landi2006}.  Their observed count rate was
$(6.61\pm2.11)$\EHM{-3}~c~s$^{-1}$ in the 0.2--10~keV band.
\citet{Landi2006} converted the count rate to 
the absorption-corrected flux of 5.4\EHM{-13}\fluxHM\ in the
2--10~keV band by assuming a Galactic column density of
2\EHM{22}~cm$^{-2}$~\citep{Dickey1990} and a Crab-like
spectrum. However, these parameters are not supported by our
Suzaku results. We converted the count rate to the flux in
the 2--10~keV band by assuming the best-fit parameters
listed in table~\ref{tbl:spec_results} with
``WebPIMMS''\footnote{http://heasarc.gsfc.nasa.gov/Tools/w3pimms.html}.
The resulting observed flux was
$(1.43\pm0.46)$\EHM{-13}~\fluxHM.  This suggests that time
variability exists in src~B, although the Suzaku flux of
src~B may have uncertainty, as described in
section~\ref{subsec:spec}.

Since src~B is in a relatively dim region in the TeV
$\gamma$-ray band, this object seems hardly related to
HESS~J1614.  However, if src~A is the X-ray counterpart of
HESS~J1614, there still may be a remote possibility that
src~B may also be physically connected to HESS~J1614, as
suggested by the best-fit column density.  One may speculate
that HESS~J1614 is an asymmetric PWN, such as
MSH~15-52~\citep{Aharonian2005c}, and src~B is a pulsar
driving the nebula. The TeV $\gamma$-ray morphology suggests
that HESS~J1614 may consist of double-sided structures, and
src~B seems to reside between them
(figure~\ref{fig:HESSvsXIS}).  The image of the TeV
$\gamma$-ray emission from MSH~15-52 is also elongated and
there is an indication that the TeV emission has peaks on
both sides of pulsar PSR~B1509$-$58.  Pulsars within nebulae
having a luminosity of $10^{33}$~\lumiHM\ are not unusual.
The spin down luminosity should be
$\sim10^{37}$~\lumiHM~\citep{Cheng2004}, then the TeV
$\gamma$-ray emission of HESS~J1614 can be driven by a few
percent of the spin-down luminosity. The extremely large
photon index of 3.6 is enigmatic, but consistent with the
X-ray photon index of
PSR~J1809$-$1917~\citep{Kargaltsev2007}, which could
conceivably associated with HESS~J1809$-$193.  However, at
present, we observe no pulsation or radio emission from
src~B, as shown in figure~\ref{fig:HESSvsXIS}.

\subsection{Src~C}

The best-fit column density suggested that src~C is a
foreground source. From the SIMBAD Astronomical Database
operated at CDS, Strasbourg,
France\footnote{http://simbad.u-strasbg.fr/simbad}, we found
a counterpart at $(l, b)=(\timeform{331.5819D},
\timeform{-00.6251D})$\footnote{$(\alpha, \delta)_{\rm J2000.0}=(\timeform{16h14m48.5s}, \timeform{-51D48'31.3''})$}
, which is a B9V star: HD~145703.  The best-fit temperature
and abundance are consistent with the X-ray observations of
other late-type B stars~\citep{Stelzer2003}.

\subsection{Soft emission in the blank region}

Within the bounds of our statistics, it is unclear whether
the emission observed in the blank region is thermal or
non-thermal. Given its soft and unabsorbed nature, this
source may not be connected to HESS~J1614 at all. Instead,
we suspect that the emission is due to spatial fluctuation
of the Galactic ridge emission, which is comprised of high
($\sim$ 7~keV) and low ($\sim$ 0.7~keV) temperature
components~\citep{Kaneda1997}.  Since the best-fit column
density is small, the soft excess emission may include the
low temperature component. The best-fit temperature is
between the high and low temperature components, and
therefore the soft excess may also include high temperature
component.  We found no evidence suggesting the presence of
soft emission in the hard band (0.5--2.0~keV) image of the
ROSAT All-Sky X-ray Survey Broad Band data (RASS3bb,
RASS-Cnt Broad).

\section{Conclusions}

The Suzaku results on the VHE $\gamma$-ray object
HESS~J1614$-$518 are summarized below.

\begin{enumerate}

\item 
Three X-ray objects, Suzaku~J1614$-$5141 (src~A),
Suzaku~J1614$-$5152 (src~B), and Suzaku~J1614$-$5148 (src~C)
are found in the Suzaku XIS field of view.

\item  
A hard extended source, src~A, is the best candidate for the
X-ray counterpart of HESS~J1614$-$518.

\item 
Src~A has a large flux ratio of $F(1-10~{\rm
TeV})/F(2-10~{\rm keV}) \sim 34$ and a relatively small
X-ray photon index of 1.7. It is difficult to explain these
features by assuming that the TeV emission is due to the
inverse-Compton scattering of the cosmic microwave
background and that the X-rays are due to synchrotron
radiation emitted by high-energy electrons.

\item 
The small photon index of src~A can be explained in terms of
non-thermal bremsstrahlung emission from the loss-flattened
electron distribution.

\item  
Src~B may not be physically related to HESS~J1614$-$518.
However, one remote possibility is a PWN scenario, in which
src~B is a pulsar driving the nebula.

\item 
Src~C is a foreground  B9V star, HD~145703.  

\item 
We found a soft excess emission near HESS~J1614$-$518, which
is characterized either by a power-law with $\Gamma=2.4$ or
a thermal plasma with $kT=2.7$~keV. This is probably not
directly related to HESS~J1614$-$518, but would nonetheless
form a part of the Galactic ridge emission.

\end{enumerate}

\bigskip
We are grateful to Profs.\ W.~Hoffman and S.~Funk for kindly
providing us with the HESS image. We also thank all Suzaku
members.  We used the ROSAT Data Archive of the
Max-Planck-Institut f\"ur extraterrestrische Physik (MPE) at
Garching, Germany.  This work is supported by the
Grant-in-Aid for the 21st Century COE "Center for Diversity
and Universality in Physics" from the Ministry of Education,
Culture, Sports, Science and Technology (MEXT) of Japan.  HM
is also supported by the MEXT, Grant-in-Aid for Young
Scientists~(B), 18740105, 2007, and by The Sumitomo
Foundation, Grant for Basic Science Research Projects,
071251, 2007. RY acknowledges support from the MEXT,
Grant-in-Aid for Young Scientists~(B), 18740153, 2007. KH
thanks the support from the MEXT, Grant-in-Aid for Young
Scientists~(B), 19740143, 2007.

\end{document}